 \documentclass[final,5p,times,twocolumn,authoryear]{elsarticle}

\usepackage{amssymb}
\usepackage{amsmath}
\usepackage{amsthm}
\usepackage{graphicx}
\usepackage{epstopdf}
\usepackage{booktabs}
\usepackage{subfig}
\journal{XXXX}

\begin{document}

\begin{frontmatter}



\title{Can complexity decrease in Congestive Heart failure?}


\author[rvt]{Sayan Mukherjee\corref{cor1}}
\author[focal]{Sanjay Kumar Palit}
\author[els]{Santo Banerjee}
\author[rvt1,rvt2]{MRK Ariffin}
\author[focal1]{Lamberto Rondoni}
\author[els1]{D.K.Bhattacharya}

\address[rvt]{Department of Mathematics, Sivanath Sastri College, Kolkata, India}
\address[focal]{Basic Sciences and Humanities Department,Calcutta Institute of Engineering and Management, Kolkata, India}
\address[els]{Institute for Mathematical Research, Universiti Putra Malaysia, Selangor, Malaysia}
\address[rvt1]{Mathematics Department, Faculty of Science, Universiti Putra Malaysia, Malaysia}
\address[rvt2]{Al-Kindi Cryptography Research Laboratory, Institute for Mathematical Research, Universiti Putra Malaysia, Malaysia}
\address[focal1]{Dipartimento di Matematica, Politecnico di Torino,Corso Duca degli Abruzzi 24, 10129 Torino, Italy}
\address[els1]{University of Calcutta, Kolkata, India}
\cortext[cor1]{Corresponding author, Email: msayan80@gmail.com}
\begin{abstract}
The complexity of a signal can be measured by the Recurrence period density entropy (RPDE) from the reconstructed phase space. We have chosen a window based RPDE method for the classification of signals, as RPDE is an average entropic measure of the whole phase space. We have observed the changes in the complexity in cardiac signals of normal healthy person (NHP) and congestive heart failure patients (CHFP). The results show that the cardiac dynamics of a healthy subject is more complex and random compare to the same for a heart failure patient, whose dynamics is more deterministic. We have constructed a general threshold to distinguish the border line between a healthy and a congestive heart failure dynamics. The results may be useful for wide range for physiological and biomedical analysis.
\end{abstract}


\end{frontmatter}


\section{ Introduction}
\label{}
Human heart reveals electrical discharges from specific localized nodes within the myocardium. These discharges propagate through the cardiac muscles and stimulate contractions in a coordinated manner in order to pump deoxygenated blood via the lungs (for oxygenation) and back into the vascular system. The physical action of human heart is therefore induced by a local periodic electrical stimulation. As a result of that, a change in potential can be measured during the cardiac cycle by electrodes which are attached to the upper torso of heart (usually both side of the heart). This recorded signal is known as electrocardiogram (ECG) [1]. A typical ECG waveform, consists of P,Q,R,S,T,U \& V major components, reveals definite pattern in the oscillation. However, different pattern can be observed due to changes of heart condition. The increasing and decreasing mechanical activities are the major causes for this complex phenomena in heart dynamics [2]. This paper studied the complexity of ECG signal of NHP and CHFP using the method of phase space analysis.

Phase space analysis is one of the most useful methods for explanation of long term dynamics. It is an abstract Euclidean space that reflects asymptotic nature of the interconnected variables which are responsible for the original dynamics [3$-$5]. The number of such variables is known as Embedding dimension [4] in which trajectory of the phase space can be flourished properly. For a continuous signal, reconstruction of phase space can be done by suitable time-delay and proper embedding dimension [6]. Suitable time-delay is generally obtained by the method of Average Mutual Information [7,8] and proper embedding dimension is obtained by method of False nearest neighbor [9$-$11]. However, different types of trajectory's movements have been observed in the phase space, viz; periodic, quasi-periodic, chaotic,etc, which can be described from Recurrence plot (RP) [12$-$16].

Recurrence plot is a diagrammatic representation of a 2D matrix, whose elements are considered $1$ if it is recurrent; otherwise it is considered $0$. $1$ and $0$ represented by black and white dots respectively in RP. RP quantifies the structure of the phase space; it does not have anything to do directly with the signals. The diagonal lines in RP are used to measure the complexity [13]. But those lines reflects the parallel movements of the trajectories, which explain the deterministic nature of the system [13]. So measuring complexity using diagonal lines in RP characterizes the degree of chaos present in the phase space dynamics. Recently, another measure of complexity with a different form of recurrence have been proposed in [17], which quantifies the presence of non-deterministic term in stochastic dynamics. So far, there is no particularly effective tool to quantify the complexity of a stochastic signal. Therefore, to investigate this type of complexity, we have implemented Recurrence period density (RPD) method of reference [18].

RPD is a probability measure which calculates density of the recurrent times. In RP, recurrent times are considered as length of white horizontal/vertical lines between each pair of recurrent points. It can be observed that, a system possesses more complex dynamics if length of the white lines varies frequently in RP. Consequently, variation of recurrent times becomes very high. Thus, probability of recurrent times reveals a proper justification to identify the complex dynamics. The quantification parameters Recurrence period density entropy (RPDE) measures complexity of such dynamics. In this concern, Shanon entropy is utilized by probability density of recurrent times. So far, the analysis of complexity was done with respect to the corresponding time series only [19$-$21], which does not predict the long-term behavior of the dynamics. So, to investigate the long-term behavior, it is important to study complexity of the phase space for more accurate prediction.

It is well known that a deterministic system is a model, which always produces the same output for a particular initial state. Any ODE system (autonomous/non-autonomous) $\frac{dX}{dt} = F(X, t)$ is deterministic by nature and can produce chaotic scenario with proper conditions. The most important point is, there are non-deterministic (random/stochastic) systems, generated by differential equations, produce chaos. For example, the additive noise term $(\xi)$ with a deterministic model $\frac{dX}{dt} = F(X, t) +\xi(t)$ can make the whole system non-deterministic [22], as the same initial condition does not give same output. In fact, this is an example of a non-deterministic system produces stochastic chaos. This kind of chaos can be observed in Ecological models, Lasers [23$-$25] and other semiconductor devices. In many cases noise can increase the nature of the complexity of the system, sometimes it may be useful to decrease chaos [26] to revert back the system to a regular state. Although, a noisy signal or a purely stochastic signal does not have any determinism and can be investigated properly using the corresponding time series, sometimes the non-deterministic outputs can also be generated by the governing equations. Noise induced (also added) chaos are very interesting to investigate, due to its rich complexity and non-deterministic nature. Recently there are research on noise induced synchronization and also chaos synchronization between stochastic models. Even the Chua system can be stochastic [27] with induced noise, studied both theoretically and experimental observations. In this article, we have proposed the dynamical complexity of some known nonlinear deterministic and stochastic systems as well as power noise by RPDE analysis. Further, by defining window Normalized Recurrence period density entropy (NRPDE), we have categorized ECG of NHP and CHFP into two classes. In fact, mean window NRPDE successfully distinguishes both types of ECG in term of complexity. Further, a proper threshold can be found in the mean window NRPDE by which we can conclude that whether a NHP becomes CHFP or not.

\section{Recurrence plot and Recurrence period density}
\label{}
For a $n$-dimensional phase space $X=\{(x_{i}):x_{i}\in\Re^{n}, i=1,2,...,N\}$, recurrence means closeness of any two points. Two points are considered close if their state vectors certainly lie in a $\epsilon$-neighborhood. Formally, two points $x_{i},x_{j}\in X, i=1,2,...,N$ are recurrent if $\|x_{i}-x_{j}\|<\epsilon$. The recurrent matrix is thus defined as
\begin{equation}
R_{i,j}=\Theta (\epsilon-\|x_{i}-x_{j}\|), i=1,2,...,N,
\label{this}
\end{equation}
where $\Theta$ is the Heaviside function, $\| .\|$ is Euclidean norm of the reconstructed phase space, and $\epsilon$ is radius of the neighborhood. RP corresponds recurrent and non-recurrent point by `1' (black dots) and `0' (white dots) respectively. Phase space is reconstructed by suitable time-delay ($\tau$) and proper embedding dimension ($m$). An $m$-dimensional reconstructed is given by
\begin{equation}
\{(u_{i},u_{i+\tau},u_{i+2\tau},....,u_{i+(m-1)\tau})\}.
\label{this}
\end{equation}

Appropriate choice of embedding delay $\tau$ plays an important role in this context. Selecting $\tau$ too small means that any trajectory lies close to the diagonal of RP and hence spurious recurrences appear. For too large $\tau$, state vectors in the embedded space fill a large cloud. In that case, recurrences will be difficult to find out. The optimum time-delay is chosen by AMI method [7,8]. AMI method measures average information shared by signal itself. For a signal $x(i),i,1,2,..,N$ and time-delay/lag ($\tau$), AMI is defined by
\begin{equation}
AMI(\tau)=\sum_{i=1}^{N-\tau}\frac{Pr(x(i),x(i+\tau))}{Pr(x(i))Pr(x(i+\tau))},
\label{this}
\end{equation}
where $Pr(.)$ denotes probability. In order to select suitable $\tau$, we used the first local minimum principle which is proposed by Fraser and Swinney in [7].

Embedding dimension reveals number of necessary independent coordinates for phase space reconstruction. In order to estimate embedding dimension, we used the method of false nearest neighbours (FNN). This method is based on the idea of False neighbour state. False neighbour state occurs for self crossing trajectory in a phase space. In fact, when a trajectory is projected to a space with too small dimension, trajectory crosses itself and the so called false neighbour states occur. As dimension of phase space increases, number of trajectory self-crossings and false neighbours decreases. It has been observed that, both disappear completely for too large dimension of the phase space. To determine FNN for a time series $x(i), i=1,2,..,N$, we used
the following criterion:
\begin{equation}
\frac{|x(i+m\tau)-x^{NN}(i+m\tau)|}{R}\geq A,
\label{this}
\end{equation}
where $x^{NN}$ represents points of nearest neighbourhood (NN). $R$ is given by $R^{2}=\frac{1}{N}\sum_{i=1}^{N} \{x(i)-<x>\}^{2}$, where $<x>=\frac{1}{N}\sum_{i=1}^{N} x(i)$. As per [9], A is so chosen that it stays around \\ $A=2$.

Now RP reveals various structure that provides information about the nature of phase space [13]. Parallel movements and trapping situation/laminar states are described by diagonal lines and vertical/horizontal lines respectively. Periodic/quasi-periodic phase space nature can be understood by the presence of only diagonal lines with equal/unequal time span. Rectangular like structure consists of diagonal line with some isolated points and vertical lines represents chaotic regime. However, more complex structure can be observed in RP for stochastic chaotic or purely stochastic system. Uniformly distributed isolated points corresponds uniformly distributed white noise. Single isolated points imply strong fluctuation exists in the process, i.e; the process is either uncorrelated or anti-correlated. Previous characteristics can also be verified by Recurrence period density.

The measure of RPDE is developed from the idea of recurrent time between the recurrent points [18]. The Recurrent times are calculated by counting non-recurrent points or white lines between two recurrent points $x_{i},x_{j}$ in the RP $R_{i,j}$. Thus, for any two recurrent points $x_{i},x_{j}\in R_{i,j}$, the recurrent time denoted by $T_{k}$, is defined as $T_{k}=(i-j)$. In fact, $T_{1}$ corresponds to the least recurrent time,$T_{2}$ corresponds to the next and so on. Now, for all points in $R_{i,j}$ a series of recurrent time interval $R(T_{k})$ is obtained as the number of occurrence of $T_{k}$. RPD $P(T_{k})$ is finally defined as the probability of $R(T_{k})$ among the sample space $\{R(T_{k})\}$. This is given by $(5)$.
\begin{equation}
P(T_{k})=\frac{R(T_{k})}{\sum_{k=1}^{T_{max}}R(T_{k})},
\label{this}
\end{equation}
where $T_{max}=max \{T_{k}\}$.

The changes of RPD as observed from RP are numerical illustrated for some commendable complex systems$-$\\Lorenz system, Logistic map and Stochastic system- first order ordinary differential equation with additive noise.  We computed $P(T_{k})$ for each system by ($3$) and observed their individual characteristics. This analysis is also done for one dimensional colored noise which is generated by $\frac{1}{f^{\beta}}$-law. The whole computational result reveals a clear conception about the pattern of probability density which changes due to the increase of complexity of the system.

For the
\begin{subequations}
Lorenz system-
\begin{align}
\frac{dx}{dt}&= a(y-x),\\
\frac{dy}{dt}&= x(r-z),\\
\frac{dz}{dt}&= xy-bz,
\end{align}
($a=10,r=28,x(0)=8,y(0)=9,z(0)=25$)
\end{subequations}
probability density is calculated for two different conditions: double periodic ($b=\frac{1}{3}$) and chaotic ($b=\frac{8}{3}$). Since most of the state vectors lie in parallel trajectories for $b=\frac{1}{3}$, variation of recurrent points are very few in number (Fig.1a). Consequently, probability density appears for few recurrent times (Fig.1b). Lorenz system ensembles chaotic regime with heteroclinic transition for $b=\frac{8}{3}$. As a result, lesser state vectors can be found in same parallel trajectory and different types of recurrent time are found in the corresponding RP (Fig.1c). Corresponding probability density is shown in Fig.1d.
\begin{figure*}[btp]
\begin{center}
    \subfloat[]{\label{fig.fig1a}\includegraphics[width=2.0in,height=1.8in,trim=0.0in 0in 0in 0in]{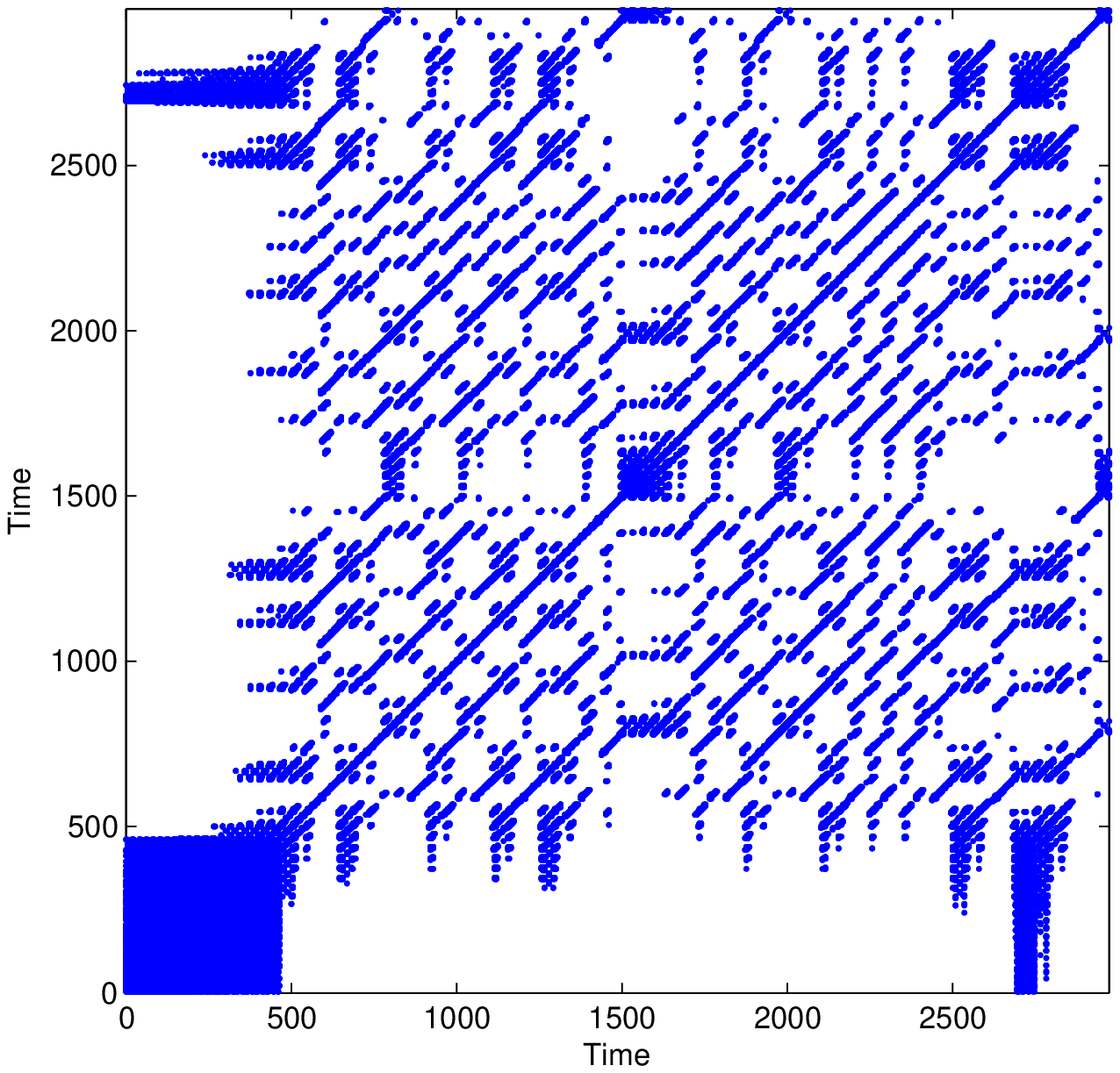}}
		\subfloat[]{\label{fig.fig1b}\includegraphics[width=2.0in,height=1.8in,trim=0.0in 0in 0in 0in]{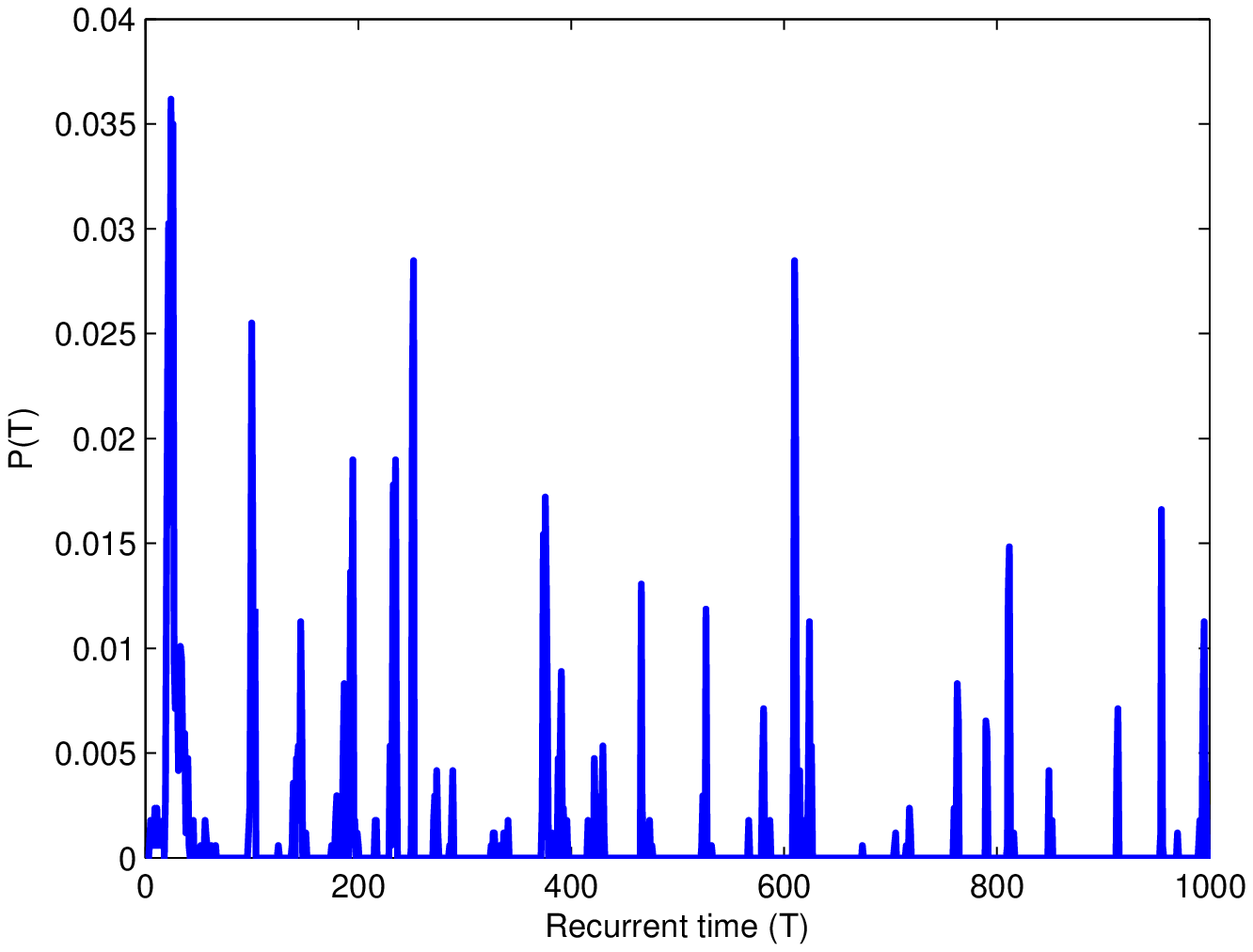}}
		
		\subfloat[]{\label{fig.fig1c}\includegraphics[width=2.0in,height=1.8in,trim=0.0in 0in 0in 0in]{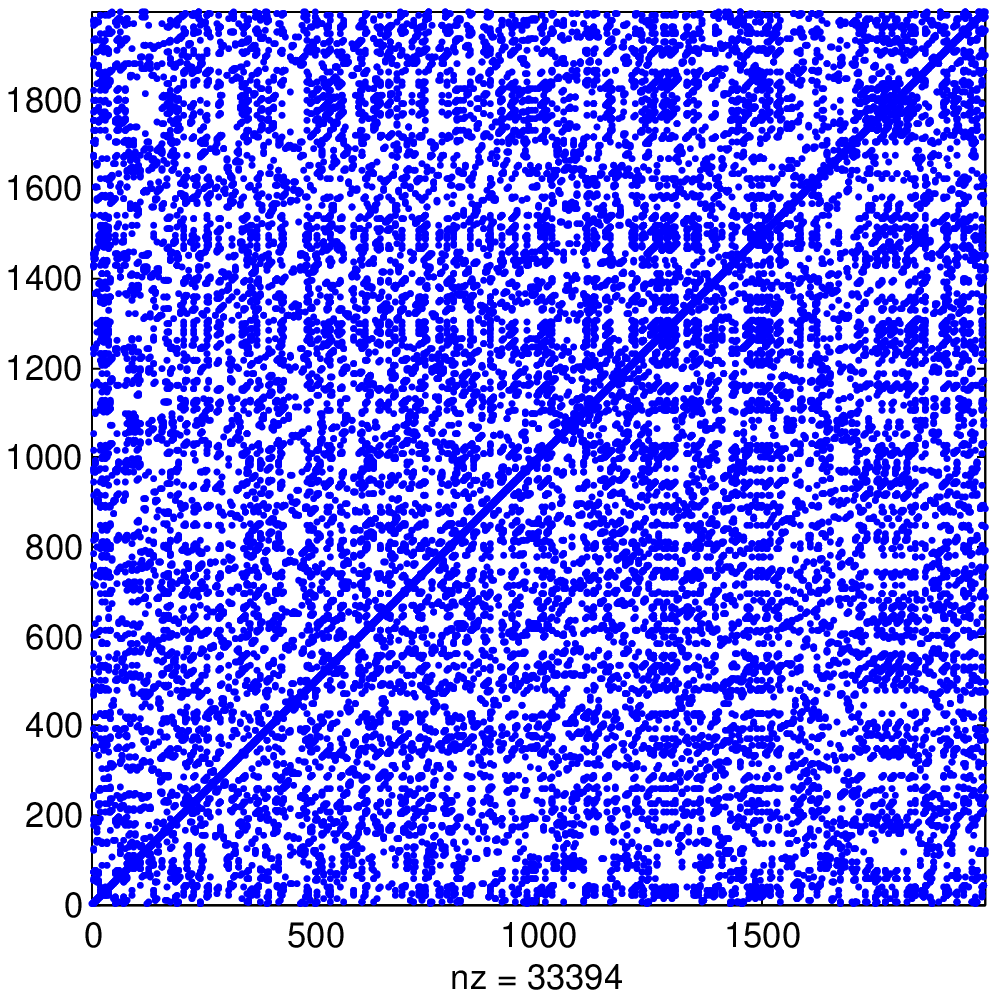}}
		\subfloat[]{\label{fig.fig1d}\includegraphics[width=2.0in,height=1.8in,trim=0.0in 0in 0in 0in]{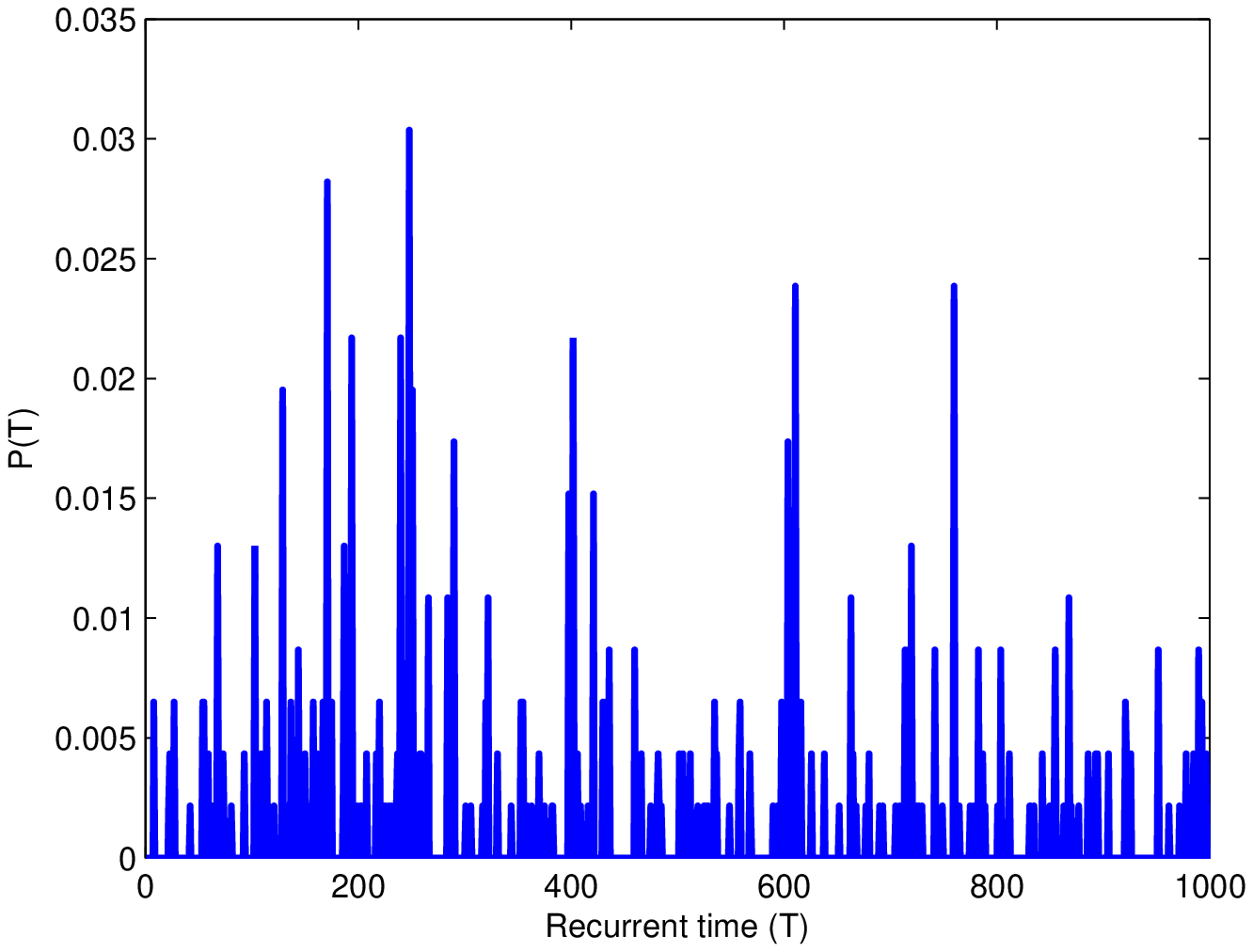}}
\end{center}
\caption{(Color online) (a) represents the RP of the system ($6$) for chaotic phase space (with $\epsilon=\frac{1}{10}D$, $D$ is the diameter of the phase space). The RPD for both $b=\frac{8}{3}$ are shown in (b) (with 1000 recurrent times) respectively. (c) represents RP of the system ($7$) for $\alpha$$=3$ ($\epsilon=0.1\sigma$, $\sigma$ is the standard deviation of the solution). The corresponding RPD are given in (d) (for $\alpha$$=3$).}
\end{figure*}

In logistic system:
\begin{equation}
x_{n+1}=\alpha x_{n}(1-x_{n}),
\label{this}
\end{equation}
we consider $\alpha=3,4$. The concept of parallel movement of trajectory is completely meaningless here. In RP, black dots represents only recurrence in $\epsilon$-neighborhood. For $\alpha=3$ and $4$, corresponding RP and probability distribution are given in Fig.1e, 1g, 1f and 1h. It is seen that, the probability of recurrent time occurs in a very short range of recurrent times (Fig.1f) for $\alpha=3$. On the other hand, various RPD appears for large range of recurrent time for $\alpha=4$ (Fig.1h). Therefore, the system ($7$) is less deterministic when $\alpha=4$.

Next consider a Stochastic differential equation (SDE) with exponential decay:
\begin{equation}
\frac{dz_{t}}{dt} = -z_{t}+\sigma(z_{t}) \zeta{(t)},
\label{this}
\end{equation}
where $\zeta{(t)}$ is additive random noise and $\sigma(z_{t})$ is strength of the noise.

For $\sigma(z_{t})=0.5$ and $1.5$, $(\tau,m)$ are found to be $(17,5)$ and $(15,4)$ respectively. The time-delay $\tau$ and embedding dimension $m$ are calculated using (3) and (4) respectively. For the purpose of unavoidable restriction the corresponding RP are not given in this manuscript.

We have also investigated the natures of RP and RPD from $\frac{1}{f^{\beta}}$-noise via phase space reconstruction ($2$). Random or disordered structure is observed in RP of the noisy signal for $\beta$$=0$(Fig.2a). In this case, the length of variation of recurrent times is higher than that of the same obtained for other values of $\beta$. On the other hand, an well patterned RP is observed in Fig.2d for $\beta$$=2$, which implies low recurrent time variation. The same investigation has also been done for the intermediate cases- $\beta$$=0$ (Fig.2b),$\beta$$=1.5$ (Fig.2c), which indicates the gradually well structured RP. Thus, study of RP on $\frac{1}{f^{\beta}}$-noise suggested that the variation of recurrent times are inversely proportional to the power of the noise $\beta$. The nature of probability density can be observed from each RPD for $\frac{1}{f^{\beta}}$-noise ($\beta=0,0.5,1.5,2$) with $T_{k}, k=1,2,..,1000$ (see Fig.2e). It is seen that, values of $P(T_{k})$ follows decreasing trend with $T_{k}, k\in [1, 1000]$ for $\frac{1}{f^{\beta}}$-noise ($\beta=0, 0.5, 1.5, 2$). In fact, the tendency of getting lower values for $P(T_{k})$ increases as $\beta$ increases (e.g; see the values of $P(T_{k})$ for green and yellow). It implies that the variation of recurrent times, i.e; $T_{k}$ decreases as $\beta$ increases. The same behaviour is observed for intermediate cases-$\beta=0.25,0.75,1,1.25,1.75$. This indicates that complexity is inversely proportional to the power noise.
\begin{figure*}[btp]
\begin{center}
    \subfloat[]{\label{fig.fig2a}\includegraphics[width=3.2in,height=2.0in,trim=0.0in 0in 0in 0in]{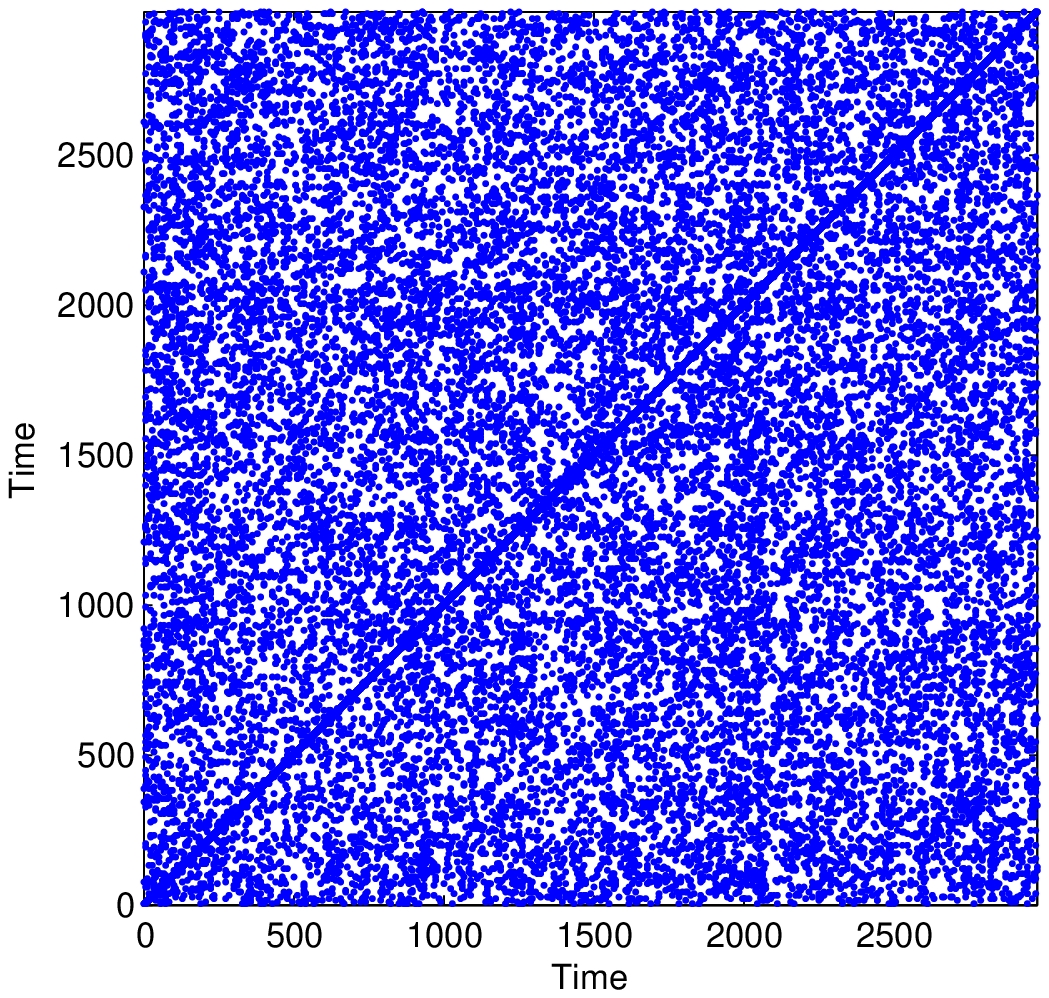}}
		\subfloat[]{\label{fig.fig2b}\includegraphics[width=3.2in,height=2.0in,trim=0.0in 0in 0in 0in]{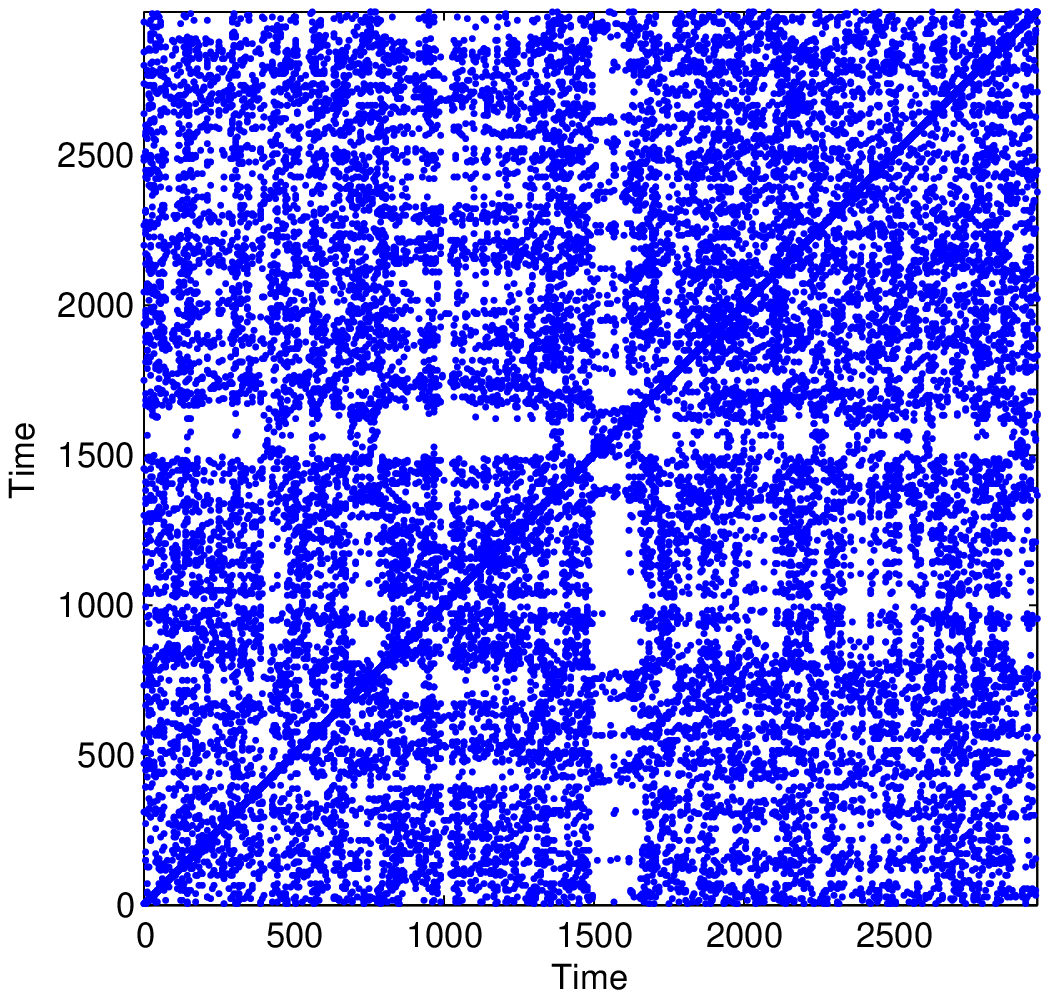}}
		
		\subfloat[]{\label{fig.fig2c}\includegraphics[width=3.2in,height=2.0in,trim=0.0in 0in 0in 0in]{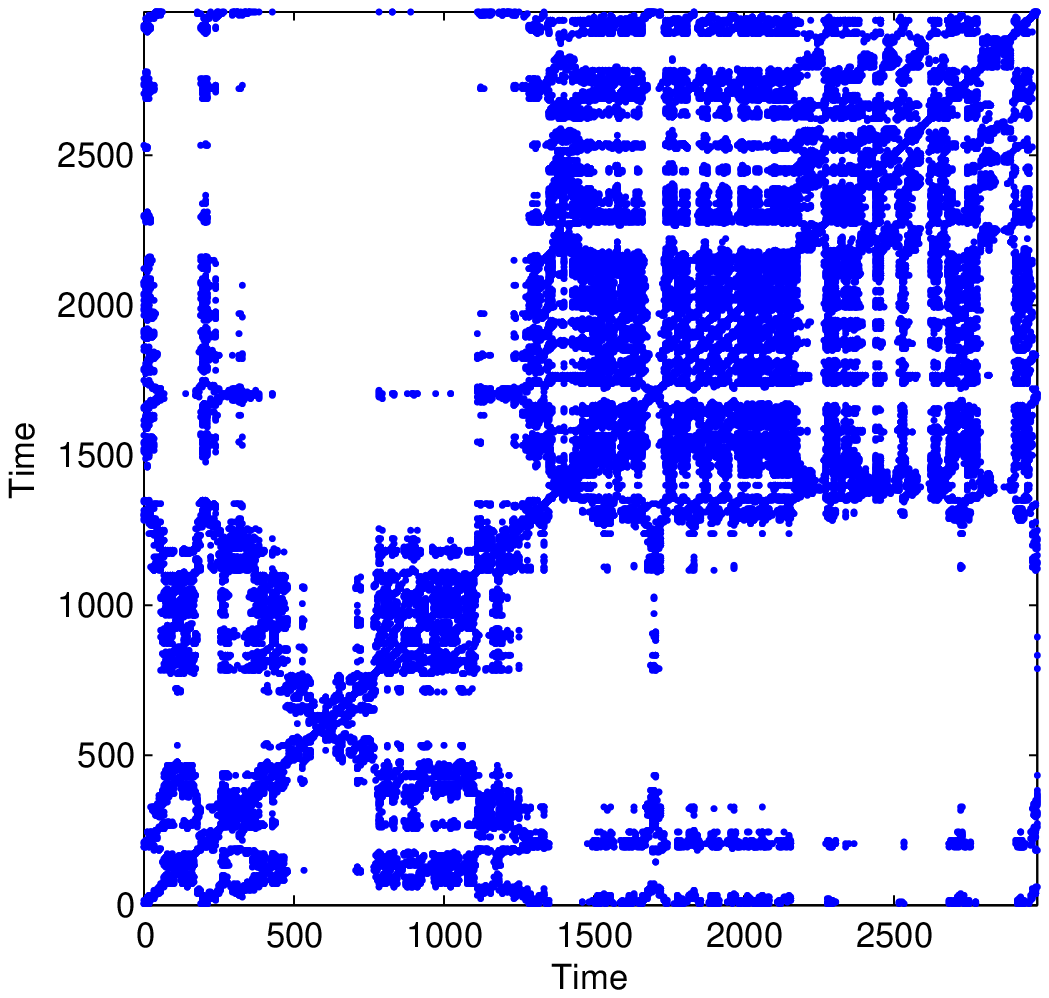}}
		\subfloat[]{\label{fig.fig2d}\includegraphics[width=3.2in,height=2.0in,trim=0.0in 0in 0in 0in]{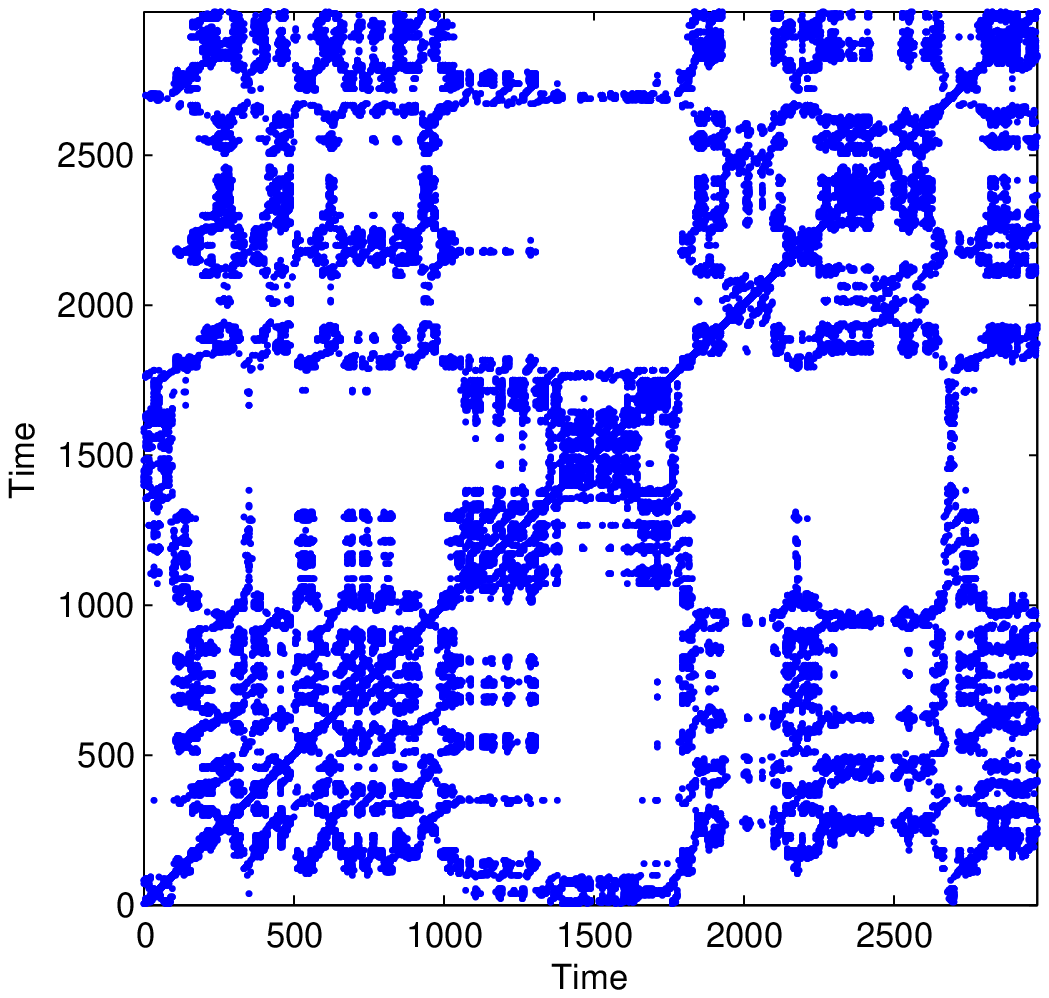}}
		
		\subfloat[]{\label{fig.fig2e}\includegraphics[width=5.2in,height=3.0in,trim=0.0in 0in 0in 0in]{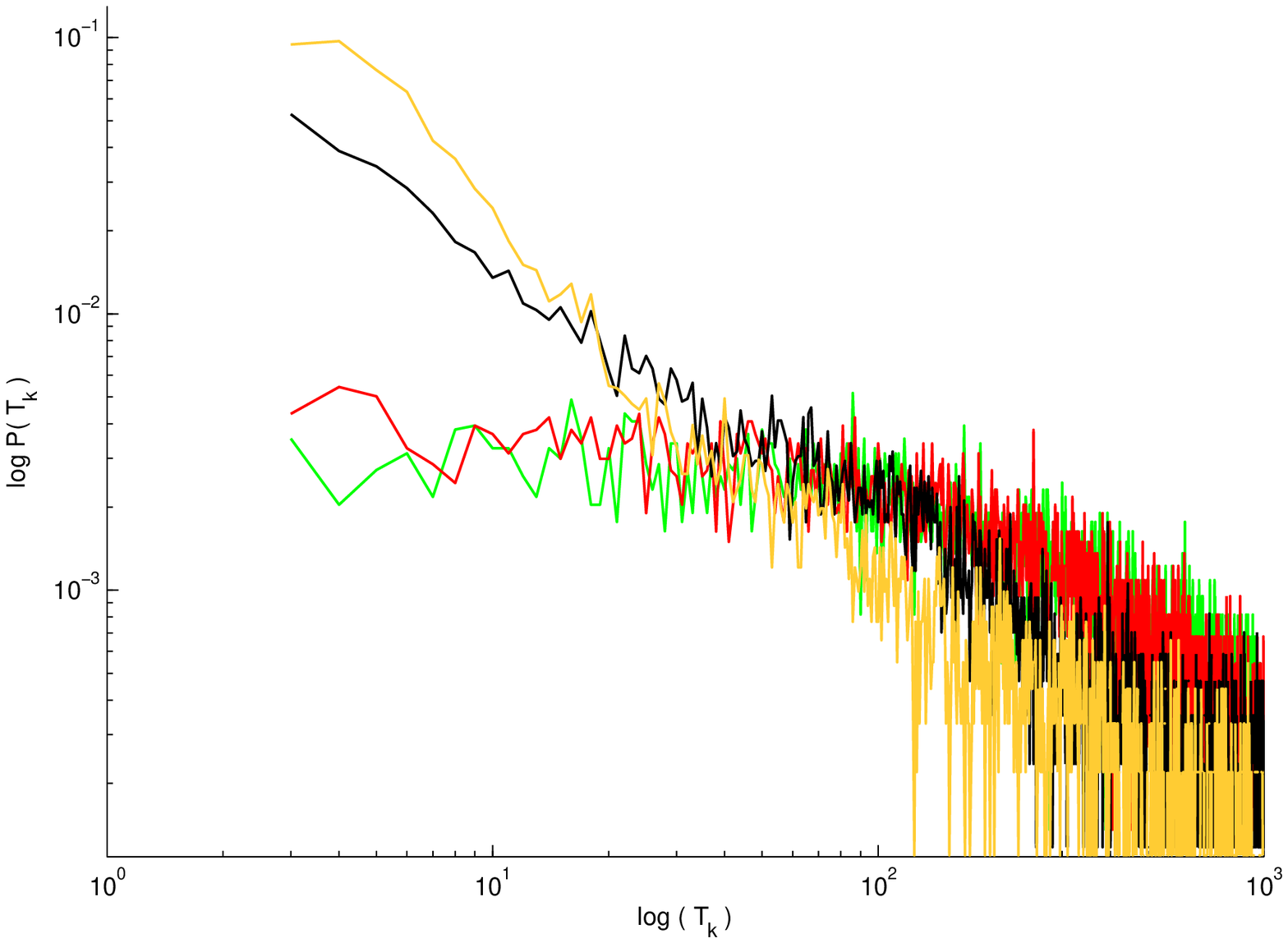}}
\end{center}
\caption{(Color online) (a)-(d) represents RP of $\frac{1}{f^{\beta}}$-noise, for $\beta$$=0,0.5,1.5,2$ respectively.(e) represents $\log-\log$ plot of RPD of $\frac{1}{f^{\beta}}$-noise, for $\beta$$=0$ (green), $0.25$, $0.50$ (red), $1.5$ (black), $2$ (yellow). Using $(2)$ and $(3)$, reconstruction parameters $(\tau,m)$ are found to be $(3,4)$ (for $\beta=0$), $(5,4)$ (for $\beta=0.5$), $(14,5)$ (for $\beta=1.5$) and $(31,5)$ (for $\beta=2$). In each case, thresholds $\epsilon$ are chosen as $\epsilon=0.1\sigma$, where $\sigma$ is the standard deviation of the power noise.}
\end{figure*}

RPD quantifies the complexity of the phase space with respect to the variation of parameters. But RPD can not measure the order of complexity. The complexity of a system can be quantified by Normalized Recurrence period density entropy (NRPDE). We have extended the idea of NRPDE to window NRPDE by taking some partitions of the solution components/signals to observe the variation of complexity within different time frames.

\subsection{Window Recurrence period density entropy}
Recurrence periodic entropy (RPDE) is based on the conception of Shanon entropy, which is defined as
\begin{equation}
H=-\sum_{i=1}^{N}p(x_{i})log(p(x_{i})),
\label{this}
\end{equation}
where $p(x_{i})$ is the probability of the event $x_{i}$ (by convention $0log0$ is taken as $0$.). Hence, RPDE is given by
\begin{equation}
H=-\sum_{k=1}^{T_{max}}P(T_{k}) \log P(T_{k}).
\label{this}
\end{equation}
Since $T_{max}$ changes with sampling time, so a normalization is necessary for RPDE. Thus, normalized RPDE is defined as
\begin{equation}
H_{norm}=-(log T_{max})^{-1}\sum_{k=1}^{T_{max}}P(T_{k}) \log P(T_{k}),
\label{this}
\end{equation}
where $\log T_{max}=-\sum_{i=1}^{T_{max}}P(i) \log P(i)$.

$H_{norm}$ actually RPDE of the reconstructed phase space, where the points are independently identically distributed.

For Window NRPDE, consider a time series $\{x(i)\}_{i=1}^{N}$ and the intervals $I_{L}=[x(1+(N-1)L),x(1+NL)], \frac{N}{L}\in \mathbb{Z}$. For each $I_{L}$, reconstruction of attractors are done by ($1$). Let $\tau_L$ and $m_{L}$ are the embedding parameters for ($1$). Then, using ($9$), we define
\begin{equation}
H_{norm}^{L}=-(log T_{max}^{L})^{-1}\sum_{T^{L}=1}^{T_{max}^{L}}P(T^{L}) log P(T^{L}),
\label{this}
\end{equation}
where $T^{L}$ and $T_{max}^{L}$ denotes the recurrent times and maximum of them over the interval $I_{L}$. $H_{norm}^{L}$ describes the normalized entropy of the phase spaces which are reconstructed over window $I_{L}$. Window NRPDE $H_{norm}^{L}$ measures variation of complexity of the dynamics over different time frame.

NRPDE is calculated for different values of the parameter for the Lorenz system, logistic map and first order SDE. The changes of the NRPD values is given in Fig.3.
\begin{figure*}[btp]
\begin{center}
    \subfloat[]{\label{fig.fig3a}\includegraphics[width=2.5in,height=2.0in,trim=0.0in 0in 0in 0in]{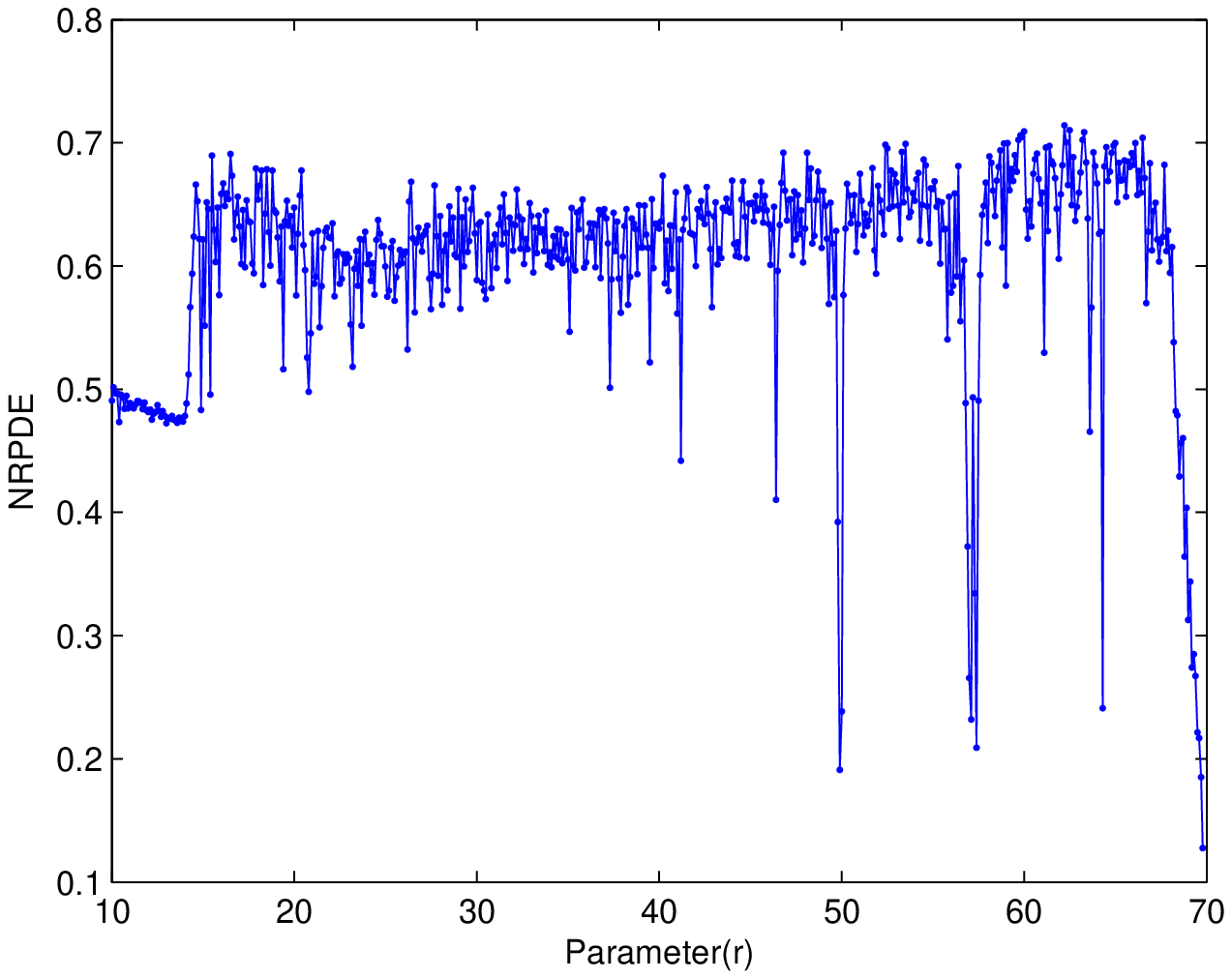}}
		\subfloat[]{\label{fig.fig3b}\includegraphics[width=2.5in,height=2.0in,trim=0.0in 0in 0in 0in]{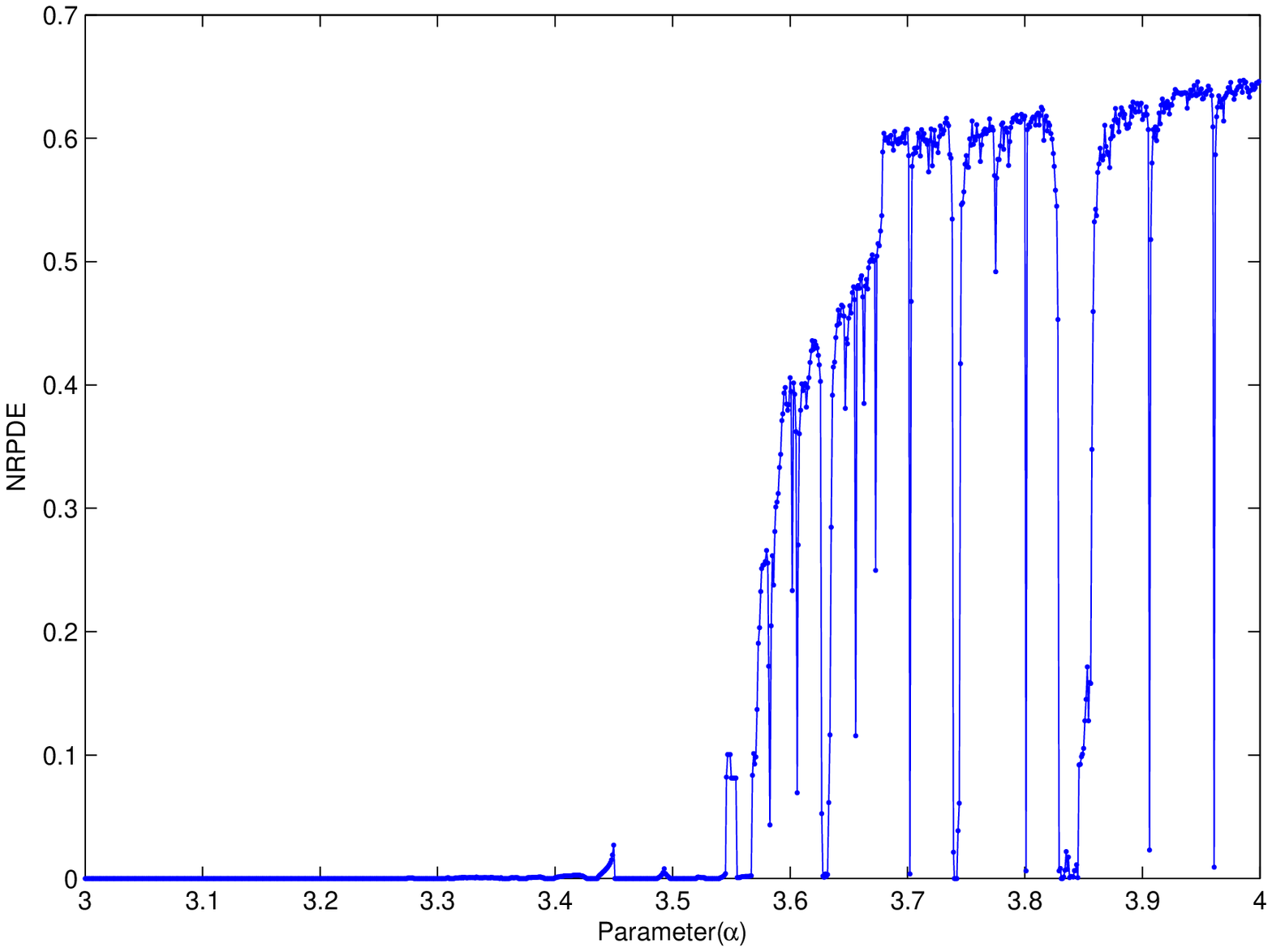}}
		\subfloat[]{\label{fig.fig3c}\includegraphics[width=2.5in,height=2.0in,trim=0.0in 0in 0in 0in]{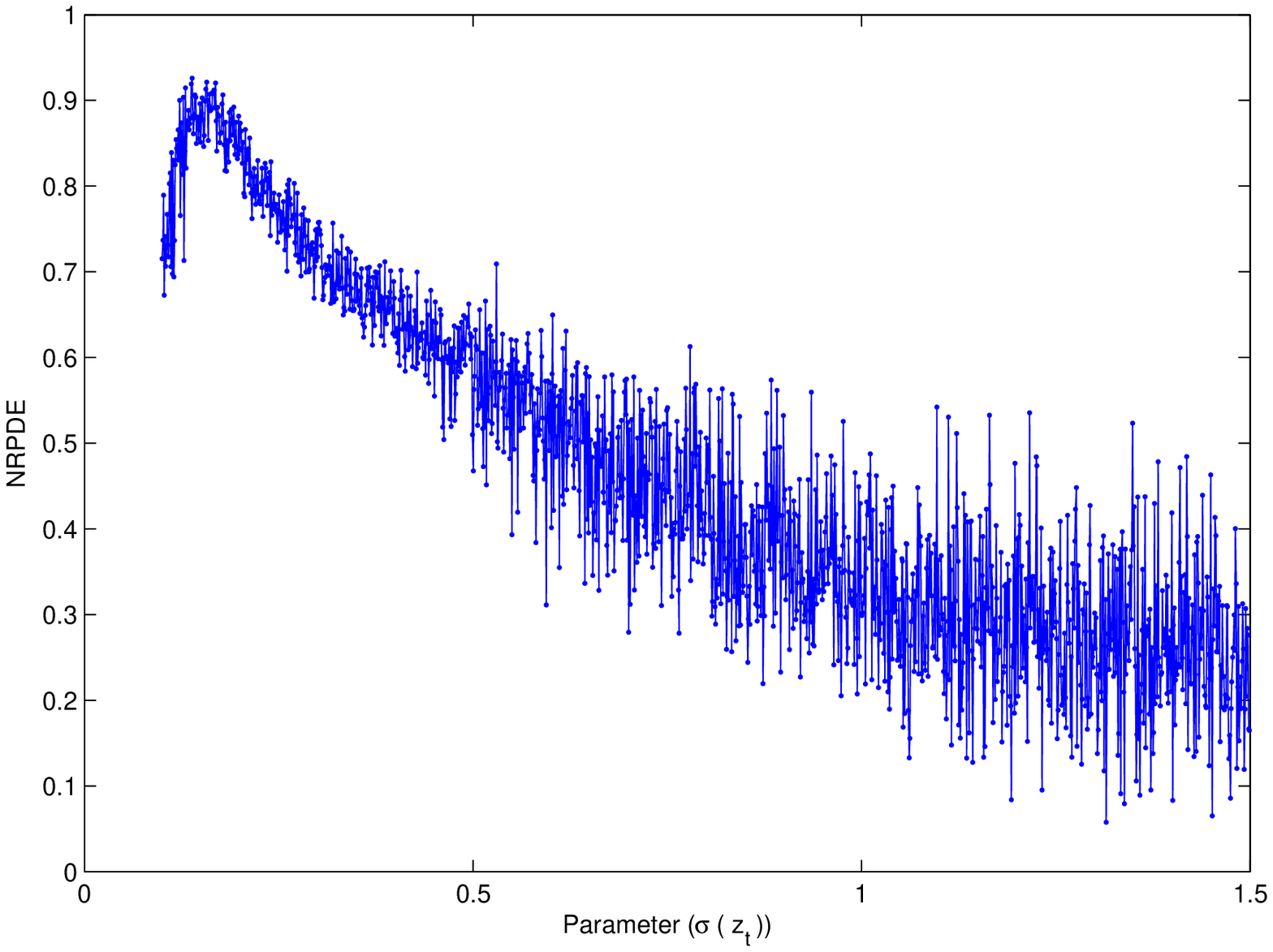}}
		
\end{center}
\caption{(Color online) NRPD of (a) system ($6$) with $r\in[11,70]$, (b) system ($7$) with $\alpha$$\in[3.001,4]$, (c) system ($8$) with $\sigma(z_{t})$$\in[0.1,1.5]$.}
\end{figure*}
It has been observed that, with the increasing values of the parameter, the corresponding NRPDE also increases (Fig.3a). Similar nature can be observed for logistic map (Fig.3b). For $\alpha$$\in [3,3.5]$, the values of NRPDE is either zero or tends to zero as evident from Fig3b . In this range, the logistic map executes periodic dynamics. Also values of NRPDE increases as $\alpha$ increases. So, changes of the nature of complexity of the phase space for Lorenz system and Logistic map can be measured by NRPD with parameters. In case of stochastic differential equations, decreasing patterns of NRPD can be observed with increment of the strength of the random noise(Fig.3c). It signifies that whenever strength of the noise increases, dynamics become less complex. Thus, NRPDE is an effective tool to describe the nature of complexity of the dynamics.

However, changes of NRPDE with parameters can not be calculated for real data. To observe the changes of dynamics complexity of a given signal, it is effective to investigate the NRPD in different time frame.

\section{Recurrence plot and Recurrence period density}
\label{}
\subsection{Application of window NRPDE on ECG signals}
\subsubsection{Collection of Cardiac signals}
Two types of ECG signals$-$($a$) Normal healthy persons (NHP), ($b$) Congestive Heart Failure Patients (CHFP) are considered as experimental subjects. All ECG signals are downloaded from Combined measurement of ECG, breathing and seismocardiogram (CEBS) database and BIDMC Congestive Heart Failure Database available in Physionet [28].

The CEBS database contains $20$ presumed healthy volunteers ($12$ men, aged $19$ to $30$, and $8$ women, aged $22$ to $28$) who were asked to be very still in supine position on a comfortable single bed and awake. This subjects are having healthy ($8$ men, $4$ women) and sedentary ($4$ men, $4$ women) life style. It is also observed that, all volunteers are non-smoker.
Signals are acquired by Biopac MP$36$ data acquisition system (Santa Barbara, CA, USA) with a bandwidth $[0.05 Hz,150 Hz]$. Each channel was sampled at 5 kHz.

The BIDMC Congestive Heart Failure Database contains $15$ long-term ECG signals (11 men, aged 22 to 71, and 4 women, aged 54 to 63) with severe congestive heart failure (NYHA class 3-4). This group of subjects was part of a larger study group receiving conventional medical therapy prior to receiving the oral inotropic agent, known as milrinone. The individual recordings are each about 20 hours in duration, and contain two ECG signals each sampled at 250 samples per second with 12-bit resolution over a range of ±10 millivolts. The original analog recordings were made at Boston's Beth Israel Hospital (now the Beth Israel Deaconess Medical Center) using ambulatory ECG recorders with a typical recording bandwidth of approximately 0.1 Hz to 40 Hz. Annotation files (with the suffix .ecg) were prepared using an automated detector and have not been corrected manually.
\subsubsection{Complexity analysis}
Complexity of NHP and CHFP can be analyzed from its corresponding reconstructed phase spaces. As a primary task, RPD is calculated on whole signals for testing complexity of the dynamics. During the reconstruction ($2$) of ECG signals, it is observed that most of the time-delay $\tau \in [20,30]$ (for NHP), $\tau \in [12,19]$ (for CHFP) and embedding dimension $m \in [3,6]$ (NHP), $m \in [5,9]$ (for CHFP) respectively. As a sample illustration, RP of the ECG signal of one NHP and one CHFP and RPD of all NHP's and CHFP's are given in Fig.4. RP of NHP (Fig.4a) shows more bowed lines than diagonal lines. Long bowed line structure indicates that the evolution of states is similar at different epochs but with different velocity, which results in rapid change of the dynamics [13]. However, these bowed lines are too short and also very few recurrent points appear in the RP of NHP. This implies that the cardiac dynamics is less deterministic and hence more complex for NHP. On the other hand, RP of the CHFP (Fig.4b) shows many diagonal lines, also many recurrent points, which reveals that the cardiac dynamics is more deterministic and hence less complex. These can be best judged from RPD analysis. It can be observed from the RPD analysis of NHP that probability density $P(T)$ appears for most of the time recurrences $T$ (Fig.4c). Thus, variable recurrent times can be observed in RP. On the other hand, $P(T)$ exists for less recurrent times (Fig.4d) for CHFP. So, recurrent times is low in the phase space and hence in the corresponding RP. Therefore, movements of the trajectories in the phase space of NHP are more complex than that of CHFP.
\begin{figure*}[btp]
\begin{center}
    \subfloat[]{\label{fig.fig4a}\includegraphics[width=3.5in,height=2.2in,trim=0.0in 0in 0in 0in]{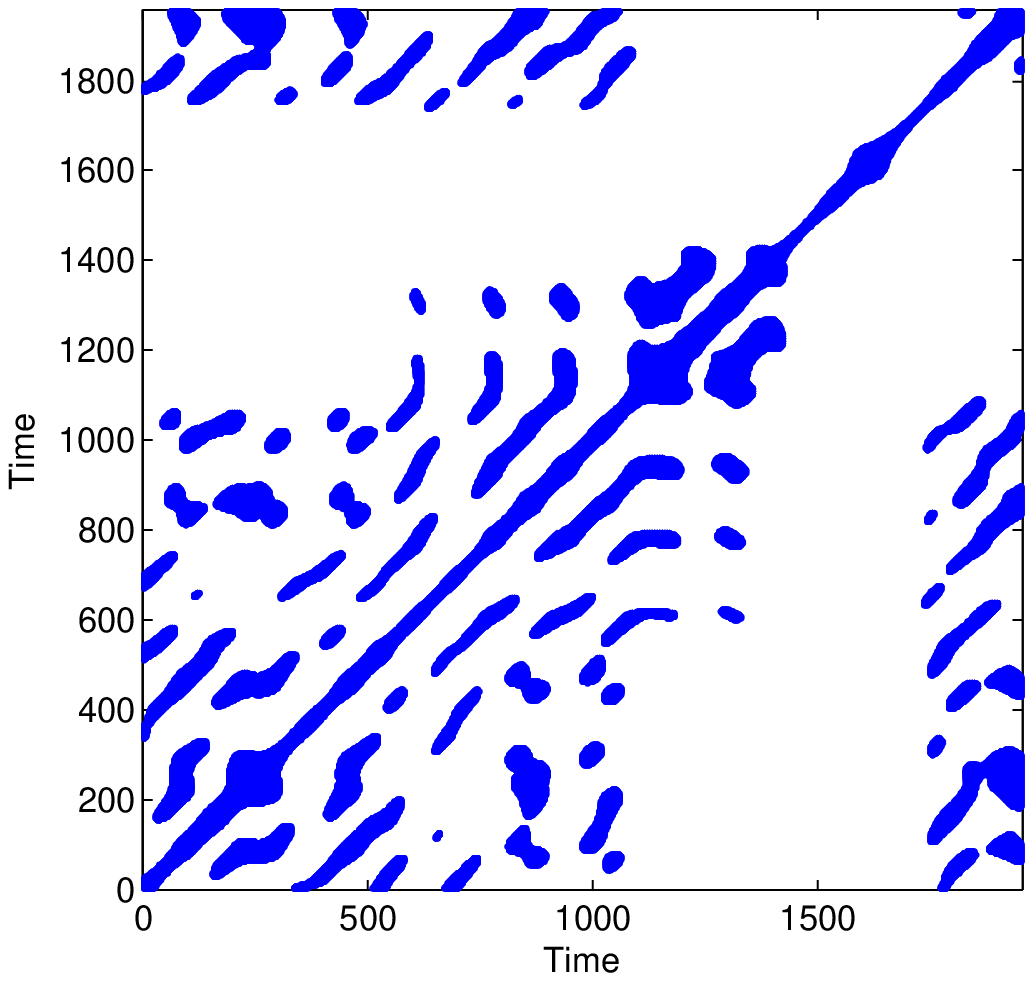}}
		\subfloat[]{\label{fig.fig4b}\includegraphics[width=3.5in,height=2.2in,trim=0.0in 0in 0in 0in]{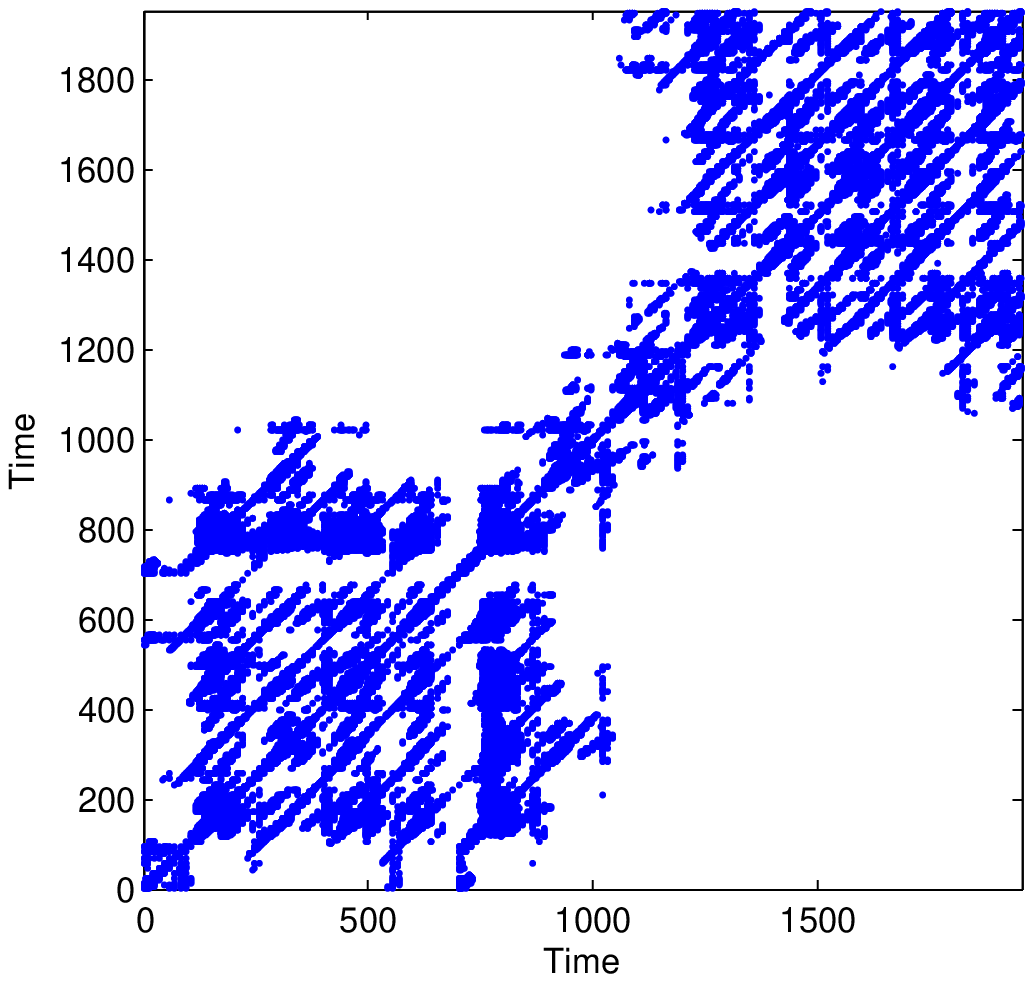}}
		
		\subfloat[]{\label{fig.fig4c}\includegraphics[width=3.5in,height=2.2in,trim=0.0in 0in 0in 0in]{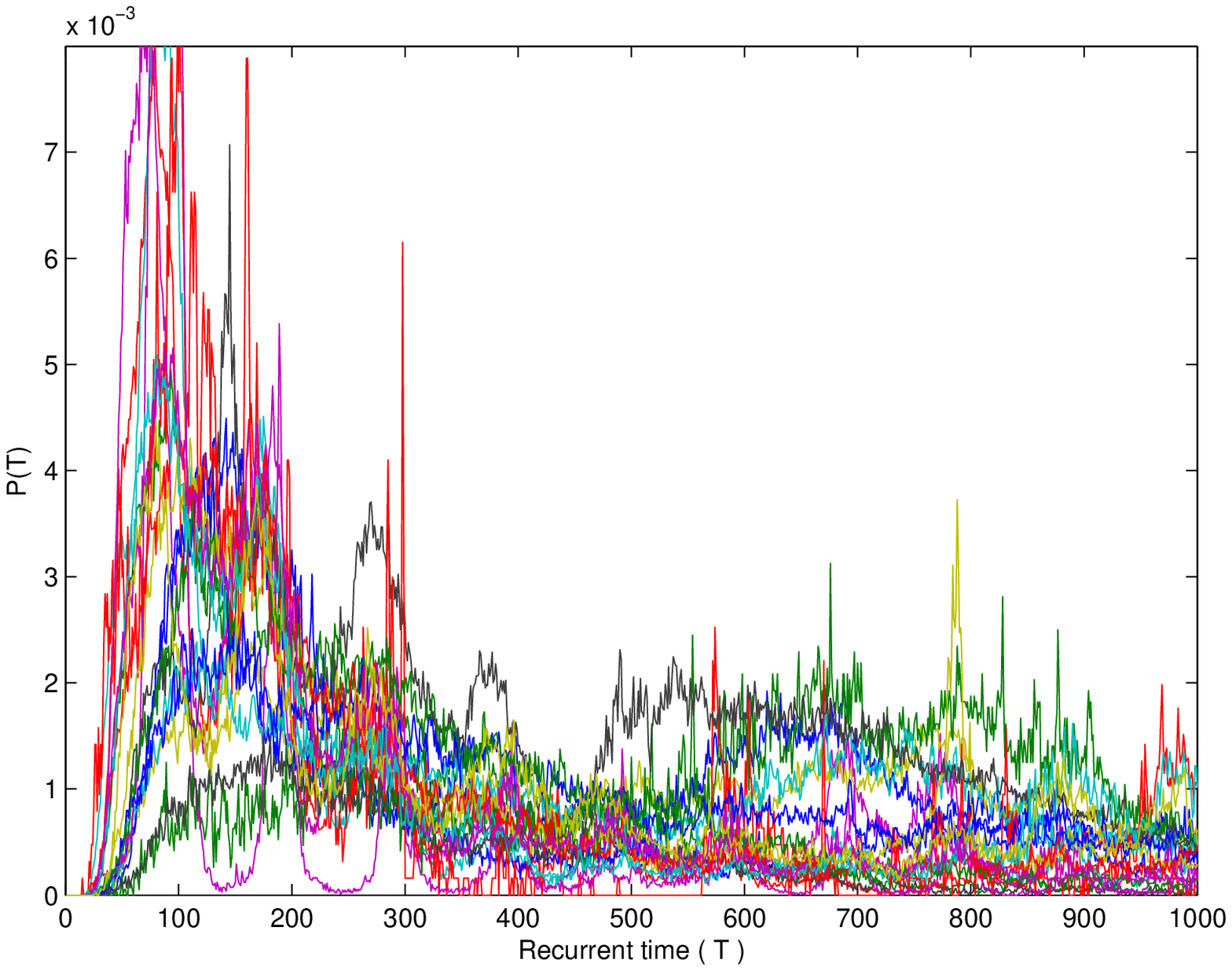}}
		\subfloat[]{\label{fig.fig4d}\includegraphics[width=3.5in,height=2.2in,trim=0.0in 0in 0in 0in]{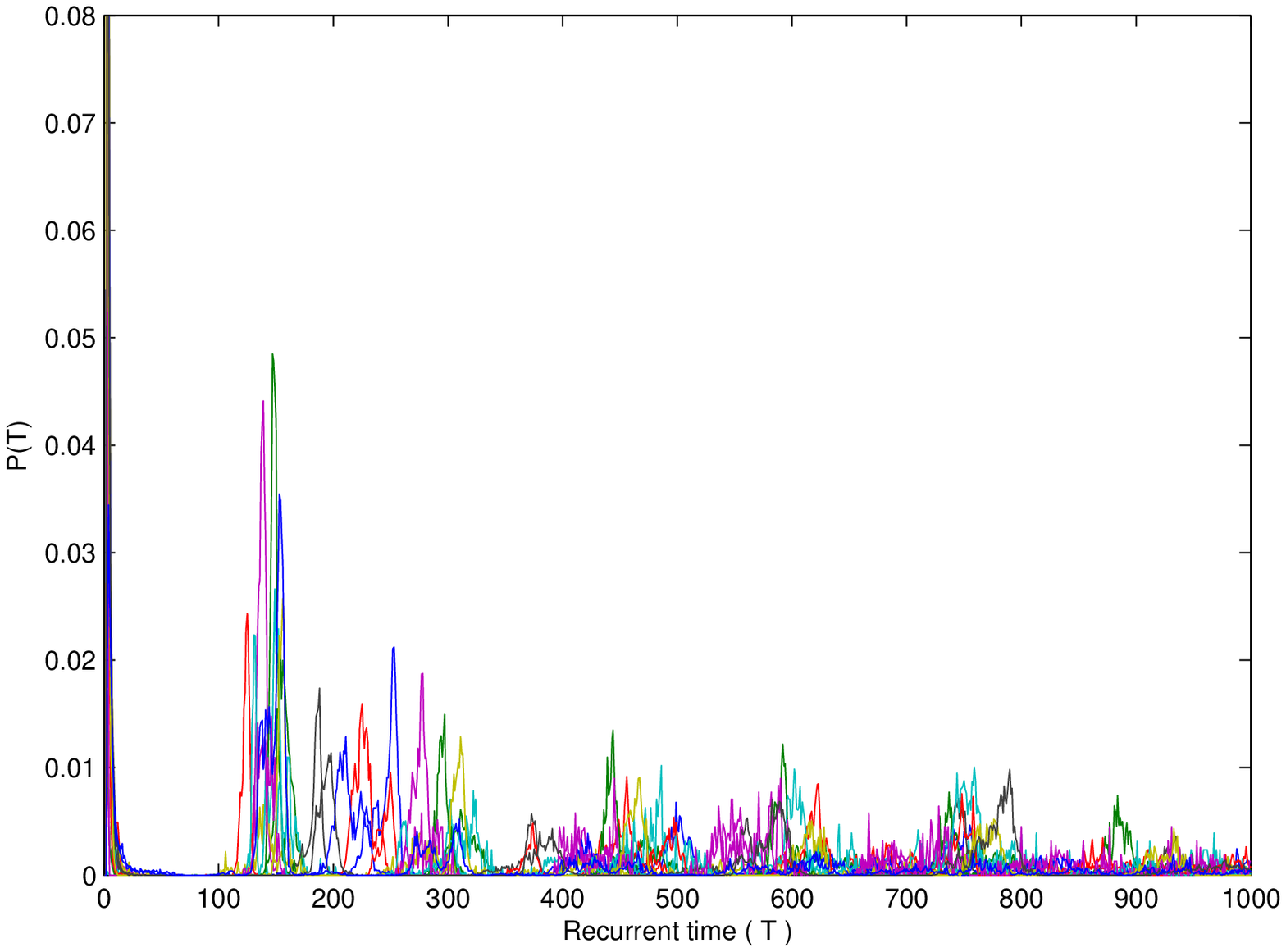}}
				
\end{center}
\caption{(Color online) (a) and (b) represents RP of one NHP (with $(\tau,m)=(20,4)$) and one CHFP signals (with $(\tau,m)=(12,6)$) respectively. In each case, thresholds $\epsilon$ are chosen as $\epsilon=0.1\sigma$, where $\sigma$ is the corresponding standard deviation of the ECG signals. RPD of (a) NHP with 1000 recurrent time, (b) CHFP with 1000 recurrent times. For NHP embedding dimension and time-delay are calculated by FNN-method and AMI-method respectively. Different colors corresponds to different signals.}
\end{figure*}
\begin{figure*}[btp]
\begin{center}
    \subfloat[]{\label{fig.fig5a}\includegraphics[width=3.85in,height=3.0in,trim=0.0in 0in 0in 0in]{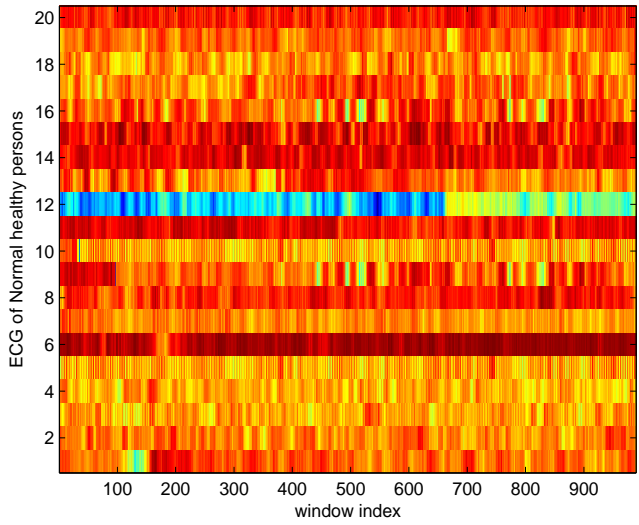}}
		\subfloat[]{\label{fig.fig5b}\includegraphics[width=3.85in,height=3.0in,trim=0.0in 0in 0in 0in]{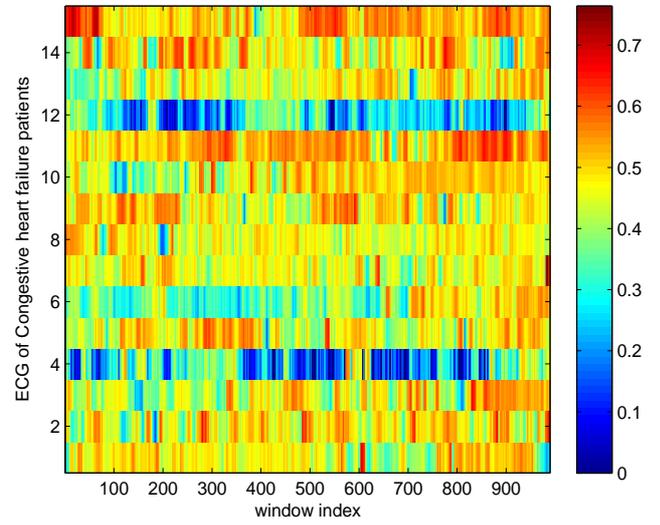}}
		
		\subfloat[]{\label{fig.fig5c}\includegraphics[width=5.85in,height=3.5in,trim=0.0in 0in 0in 0in]{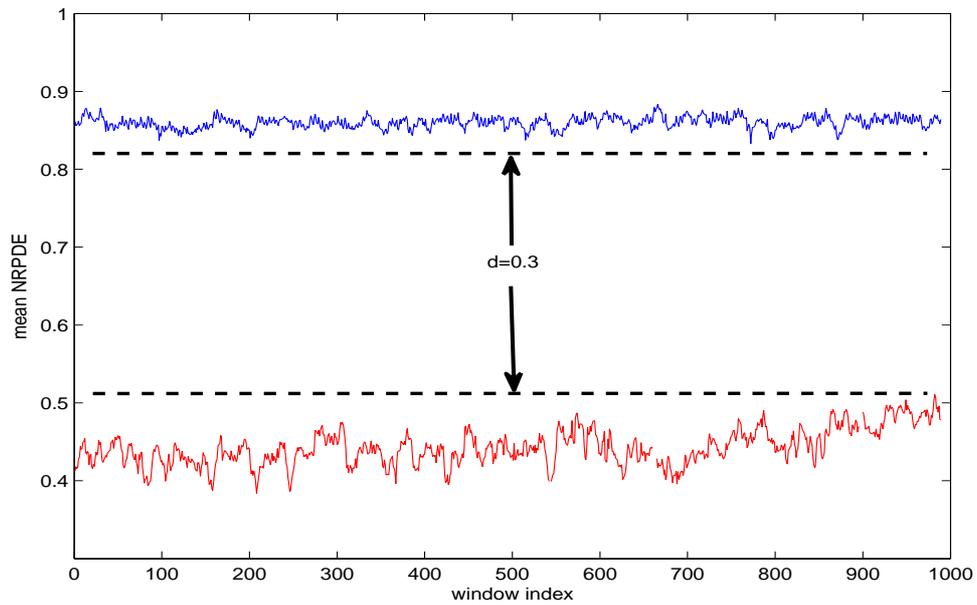}}		
				
\end{center}
\caption{(Color online) Matrix plot represents values of window NRPDE for (a) NHP, (b) CHFP. x-axis represents window index and y-axis represents the samples. In each case, $1000$ window is considered with length $10000$. (c) mean NRPDE are drawn for NHP (blue) and CHFP (red). Each color in the matrix plot represents values of NRPD of each signals with the corresponding windows.}
\end{figure*}
The corresponding window NRPDE is given in Fig.5. In Fig.5a, window NRPDE for NHP shows always higher values (see color bar). The higher values of NRPDE corresponds to higher complexity in the dynamics. So, window NRPDE infer that ECG signals of NHP groups always possess higher complexity in the dynamics. It is also observed that, window NRPDE for CHFP is less than the same for NHP (Fig.5b). This implies that, there exists less complex dynamics in the ECG signals of CHFP. Thus it is important to find a threshold to distinguish NHP and CHFP. The mean of window NRPDE is an effective tool in this context. In Fig.5c, we have shown the changes of mean window NRPDE with respect to the window index. It is observed that, the values of mean window NRPDE in all cases of NHP lie above $0.82$ and thus whenever mean window NRPDE comes below $0.82$, the cardiac dynamics tending to be less complex in nature which may lead to congestive heart failure (Fig.5c). On the other hand, the values of mean window NRPDE is less than $0.52$ in all cases of CHFP. So whenever mean window NRPDE is found to be greater than $0.52$ for some CHFP, the cardiac dynamics becomes more complex in nature which corresponds to healthy cardiac dynamics. From the clinical perspective, this may indicate the recovery of the patient from congestive heart failure.  It is also observed that the the difference between the mean window NRPDE of NHP and CHFP ($d$)  maintains a fixed distance of $0.3$ in all cases. All these results well correlate with clinical and experimental biomedical observations [2]. In this context, it may be noted that the similar observations for time series of consecutive heartbeat intervals for NHP and CHFP have also been observed in case of multiscale entropy (MSE) method [29,30]. However, the present method of window NRPDE is obtained from the phase space of the given signal, while MSE is defined for the signal itself. As we know that the phase space gives the long term behaviour of the signal, the window NRPDE method is more robust than MSE that depends on the short term behaviour of the signal. Thus it is always expected that window NRPDE method will produce more unbiased result in the classification of NHP and CHFP.
\section{Conclusions}
In this article we have studied the nature of complexity of ECG signals for normal healthy persons (NHP) and congestive heart failure patients. We have collected data from Physionet signal archive for 20 normal persons and 15 congestive heart failure patient (CHFP). The intention is to identify and distinguish normal healthy persons and congestive heart failure patients in terms of complexity. Complexity is recognized from RP of the corresponding phase spaces. In fact, probability density of recurrent times, i.e; RPD have been calculated from RP, to observe an overall view of complexity of the dynamics. The order of complexity have been measured by NRPDE, the corresponding dynamics was verified well with known chaotic models, stochastic systems and power noise. It has been observed that the dynamics of ECG for a NHP is more complex in nature compared to the same for a CHFP. It is also observed that a congestive ECG follows a deterministic nature whereas the dynamics of a healthy ECG is random. Finally a threshold value in the mean window NRPDE has been found out to distinguish the border line between NHP and CHFP. Though further investigation into the physiological changes in ECG of NHP and CHFP is needed to consolidate the present study, our work reveals an alternative and at the same time a befitting procedure for the classifications of ECG signals of NHP and CHFP. Possible uses of this tool in clinical settings include the early detection of the life threatening congestive heart failure.

\end{document}